\documentclass{article}
\usepackage{spconf,amsmath,graphicx}

\title{Image reconstruction enhancement via masked regularization}
\name{Victor Churchill, Anne Gelb\thanks{This work is supported in part by the grants NSF-DMS 1502640, NSF-DMS 1732434, and AFOSR FA9550-18-1-0316.}}
\address{Department of Mathematics, Dartmouth College}
\begin{document}
\topmargin=0mm
%
\maketitle
\begin{abstract}
Image reconstruction based on an edge-sparsity assumption has become popular in recent years. Many methods of this type are capable of reconstructing nearly perfect edge-sparse images using limited data. In this paper, we present a method to improve the accuracy of a suboptimal image resulting from an edge-sparsity image reconstruction method when compressed sensing or empirical data requirements are not met. The method begins with an edge detection from an initial edge-sparsity based reconstruction. From this edge map, a mask matrix is created which allows us to regularize exclusively in regions away from edges. By accounting for the spatial distribution of the sparsity, our method preserves edge information and and furthermore enhances suboptimal reconstructions to be nearly perfect from fewer data than needed by the initial method. We present results for two phantom images using a variety of initial reconstruction methods.
\end{abstract}
\section{Introduction}
\label{sec:intro}
A goal in the imaging science community is to be able to reconstruct images from a small amount of data. Compressed sensing algorithms, e.g. \cite{candes2006robust}, use edge-sparsity based reconstruction methods to accomplish this task. Theoretical exact reconstruction guarantees exist given particular conditions on the forward model, data collection pattern, amount of data, and edge sparsity of the image. There are also empirical results showing the amount of data required to achieve near-perfect reconstructions for specific images. This paper is concerned with when these requirements are \emph{not} met, specifically when too few data are used and a suboptimal image is returned. While the intensity values in images created from too few data using edge-sparsity based reconstruction methods may not be ideal, in many cases the edge locations in the image are faithful to those of the ground truth image. In this paper, we present an algorithm which demonstrates that if the edge locations of the reconstruction are accurate ``enough'', it is possible to improve the suboptimal reconstruction recovered from limited data.

The algorithm presented is based on the edge-adaptive $\ell_2$ regularization method from \cite{churchill2018edge} for signal and image reconstruction from (non-uniform) Fourier data, which used a pre-processing $\ell_1$ regularization based edge detection method to extract edges before applying $\ell_2$-regularized reconstruction. Specifically, an edge mask was generated so that the $\ell_2$ regularization would only occur in smooth regions of the image. Further theoretical and empirical support for this two-stage image reconstruction was presented in \cite{churchill2019edge}, where it was shown that given a perfect mask of edge locations, minimizing the ``edge-masked'' cost function will perfectly reconstruct the image. In this case, nearly perfect reconstruction was empirically shown to be possible using only a single radial line through the 2D data collection space for the application of computed tomography (CT).

Here we assume we are given an image that has been reconstructed using an edge-sparsity based method. Note that the data for this image can be acquired in a multitude of ways. The proposed algorithm has two steps. The first step is creating a mask that gives edge locations. To achieve this, an edge transform is applied to the given image data and the result is thresholded to determine the approximate edge locations. This mask is then used in a second reconstruction step. Using the same acquired data that the initial reconstruction method used, an edge-masked $\ell_2$-regularized reconstruction is performed. The mask allows the method to regularize away from edges, which has been shown to improve accuracy. When broken down into its component steps, this post-processing enhancement technique is in fact an edge-masked image reconstruction method that is informed by an initial image reconstruction with fairly accurate edge locations.

Note that while there are similarities in the goal of our proposed method and iteratively reweighted or edge guided image reconstruction methods \cite{candes2008enhancing,chartrand2008iteratively,guo2010edgecs,guo2012edge}, i.e. regularizing away from edges in order to account for the spatial distribution of the sparsity, our algorithm is not intended to compete with these other methods. To the contrary, our technique functions as a post-processing step to further enhance an image created with one of these other methods, and uses fewer data than typically required for an ideal reconstruction. This accuracy improvement comes relatively cheaply at the computational cost of a single $\ell_2$-regularized minimization. In what follows we show that our new algorithm has the potential to improve reconstructions from a variety of edge-preserving reconstruction methods using two different edge-sparse phantoms in experiments when (i) data requirements for near-perfect reconstruction are not met and (ii) zero-mean Gaussian noise is added to the data.

\begin{figure}[t]
\centering
\includegraphics[width=4.0cm]{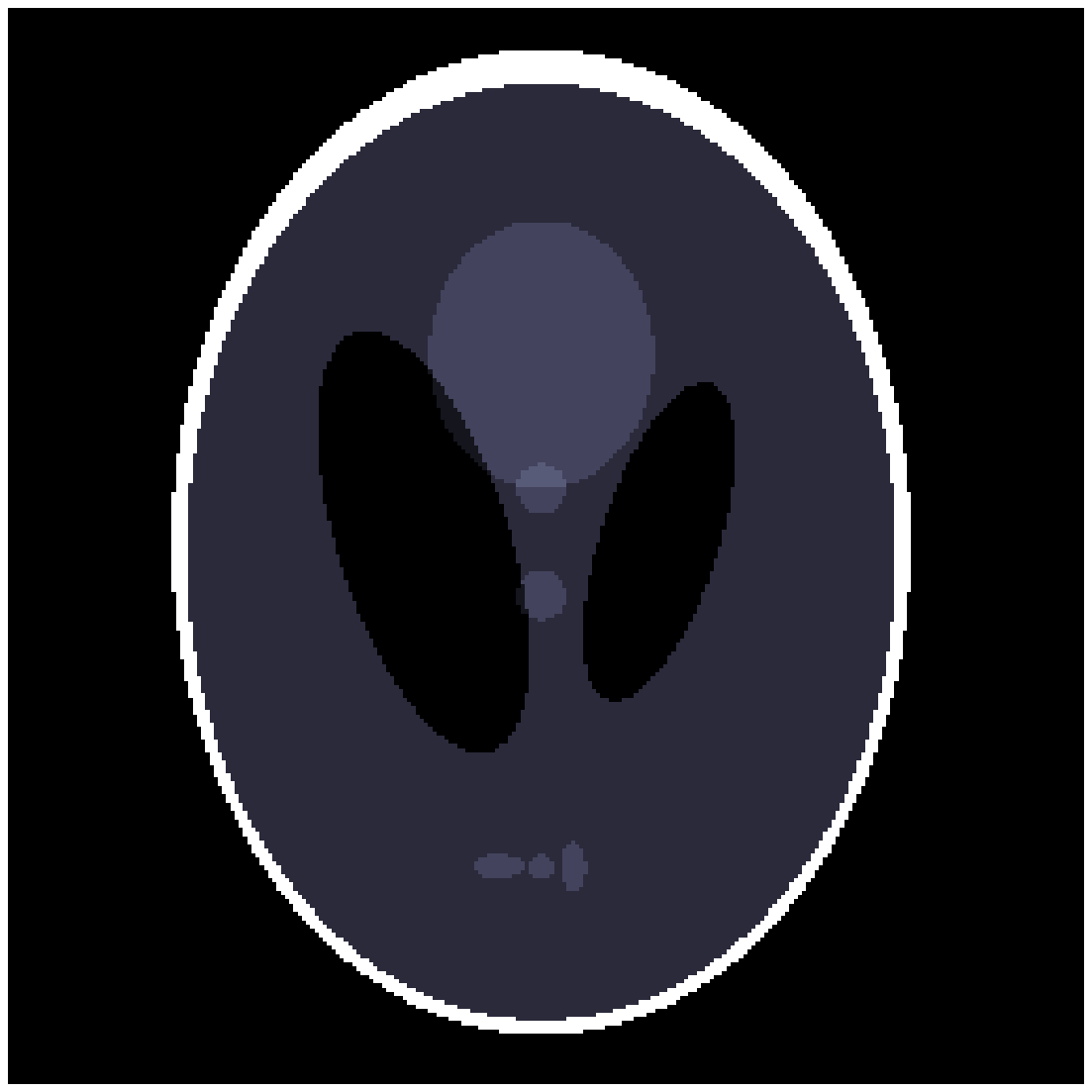}
\includegraphics[width=4.0cm]{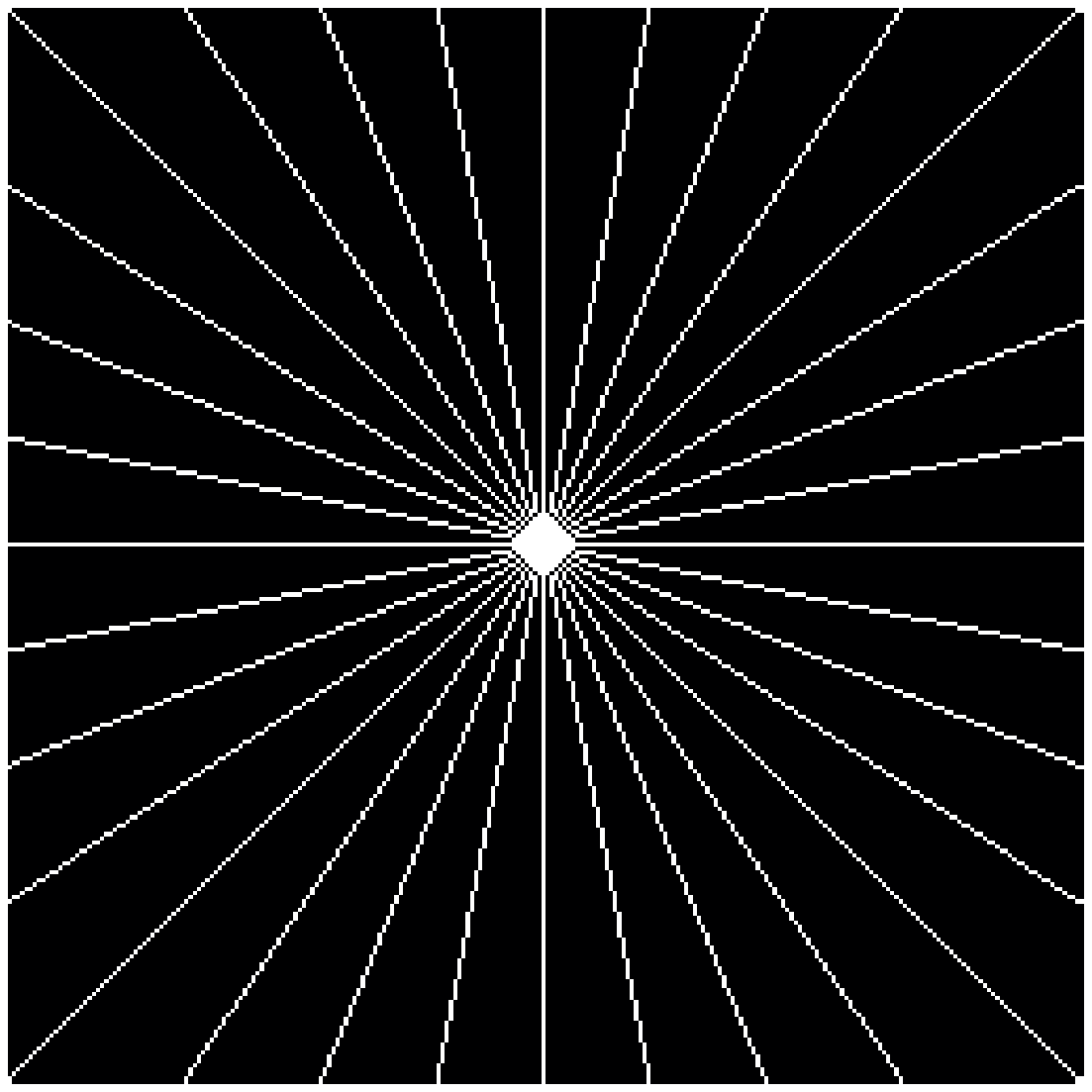}
\includegraphics[width=4.0cm]{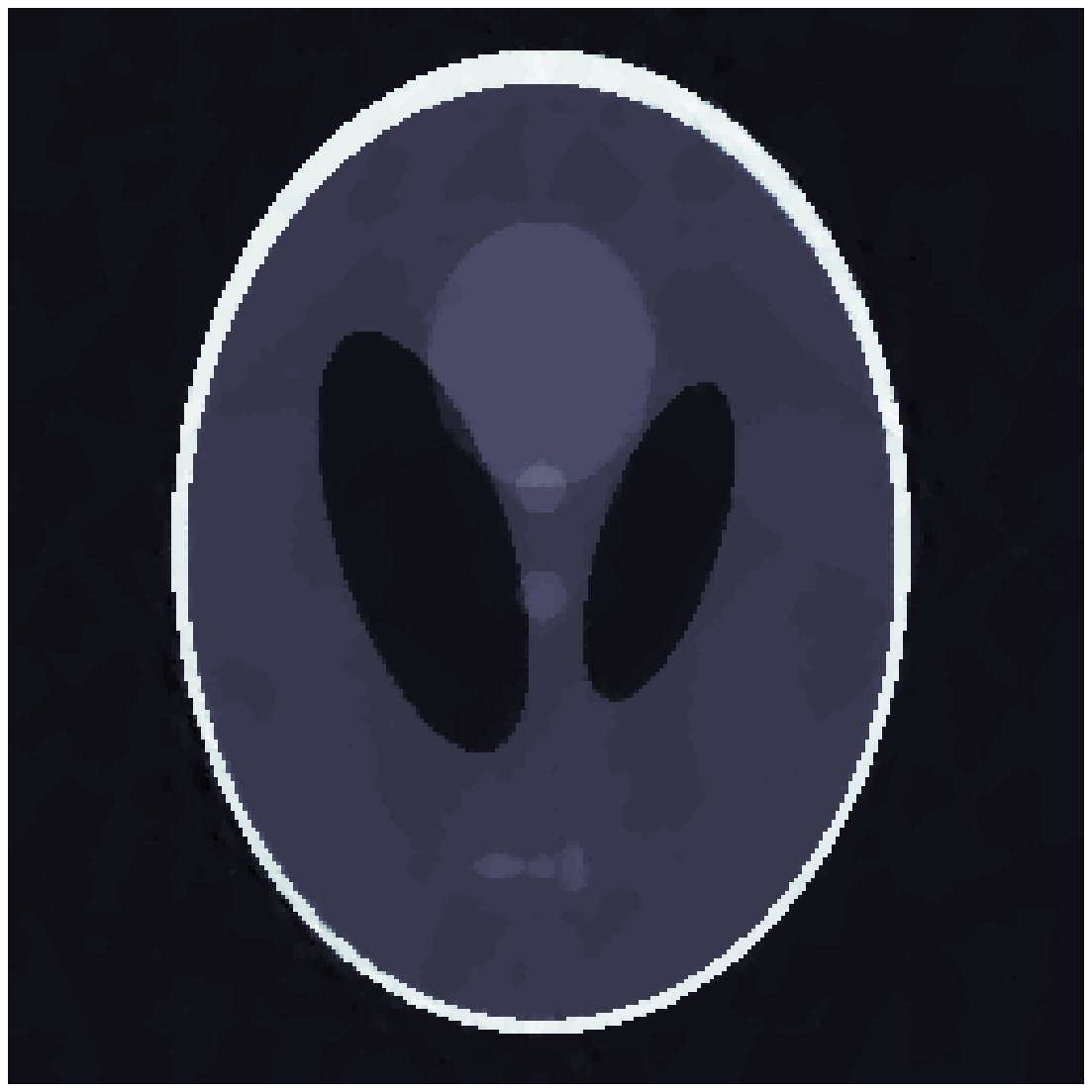}
\includegraphics[width=4.0cm]{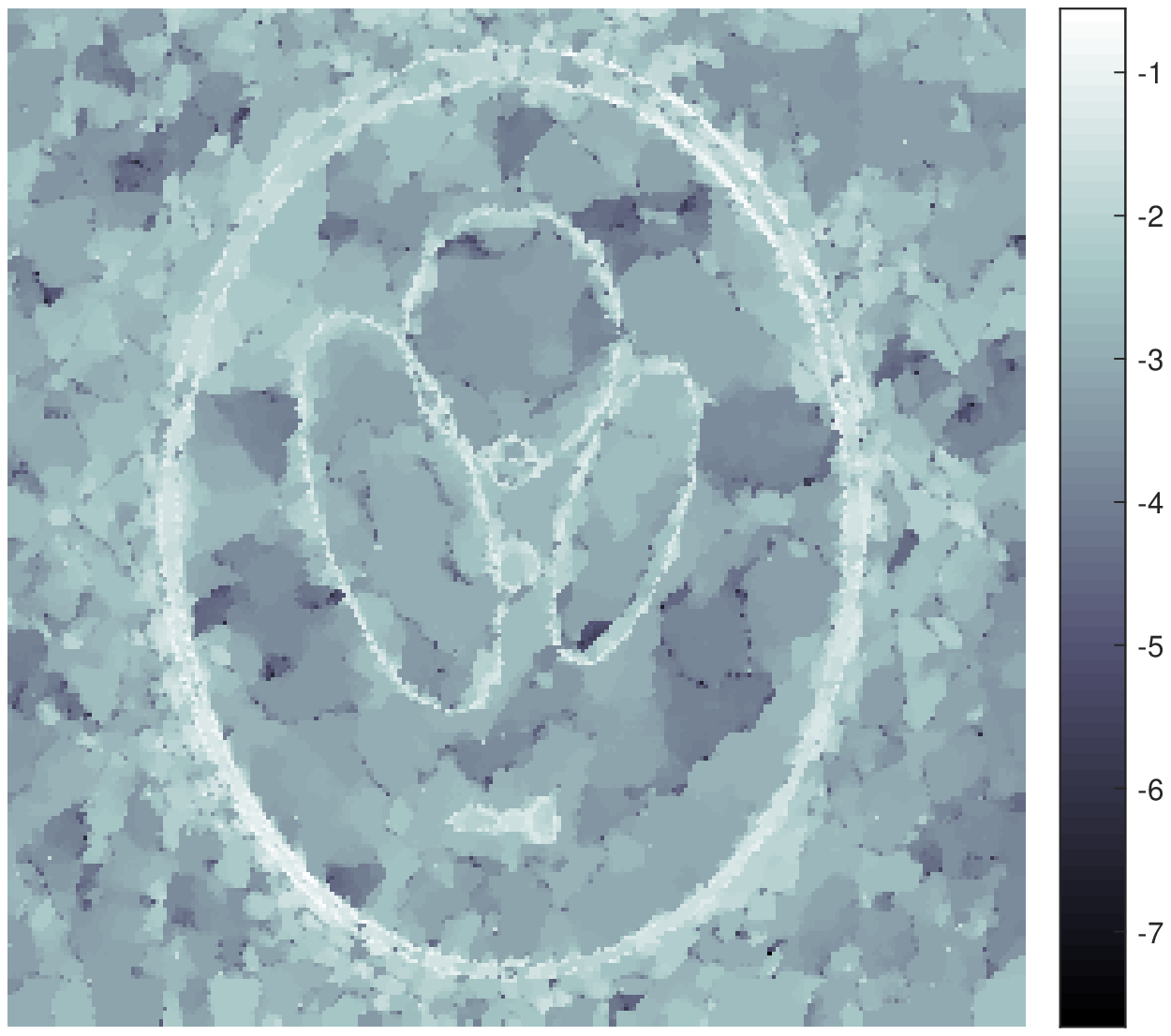}
\caption{Image reconstruction from 16 radial lines. (top left) Shepp-Logan phantom. (top right) Fourier sampling domain. (bottom left) TV-regularized reconstruction via Eq. (\ref{eq:TV}). (bottom right) point-wise error plot.}
\label{fig:TV}
\end{figure}

\section{Algorithm}\label{sec:algorithm}
This section explains the edge-masked image reconstruction enhancement algorithm through an illustrative example. Consider an image that has already been reconstructed using an edge-preserving reconstruction method, for example the isotropic total variation (TV) regularization technique, \cite{rudin1992nonlinear}. In the noise-less form, this method solves
\begin{align}\label{eq:TV}
\begin{split}
\arg\min_\mathbf{x} TV(\mathbf{x})\quad\text{subject to}\quad \mathbf{A}\mathbf{x}=\mathbf{b},
\end{split}
\end{align}
where $\mathbf{A}$ is the forward model, $\mathbf{b}$ is the data collected, and
\begin{align}
TV(\mathbf{x}) = \sum_{i,j} \sqrt{|\mathbf{x}_{i+1,j}-\mathbf{x}_{i,j}|^2+|\mathbf{x}_{i,j+1}-\mathbf{x}_{i,j}|^2}.
\end{align}
When noise is present, Eq. (\ref{eq:TV}) is modified to
\begin{align}
\arg\min_\mathbf{x} ||\mathbf{Ax}-\mathbf{b}||_2^2+\lambda\cdot TV(\mathbf{x}),
\end{align}
where $\lambda>0$ is the user-defined regularization parameter that balances noise reduction, fidelity, and edge sparsity. Note here that $\mathbf{x}$ is an $N\times N$ image. This example considers image reconstruction from radially-sampled discrete Fourier coefficients, where $\mathbf{A}=\mathbf{F}$, the 2D discrete Fourier transform, and $\mathbf{b}=\mathbf{\hat{f}}$, the 2D discrete Fourier coefficients of the ground truth image. We note that other forward models can also be accommodated. In \cite{candes2006robust} and \cite{candes2008enhancing}, it is shown that Eq. (\ref{eq:TV}) was capable of near-perfect reconstruction of the Shepp-Logan phantom, \cite{shepp1974fourier}, from measurements collected on 17 radial lines of 2D Fourier space. Figure \ref{fig:TV} shows the result of Eq. (\ref{eq:TV}) using measurements collected on only 16 radial lines instead of 17.

\begin{figure}[t]
\centering
\includegraphics[width=4.0cm]{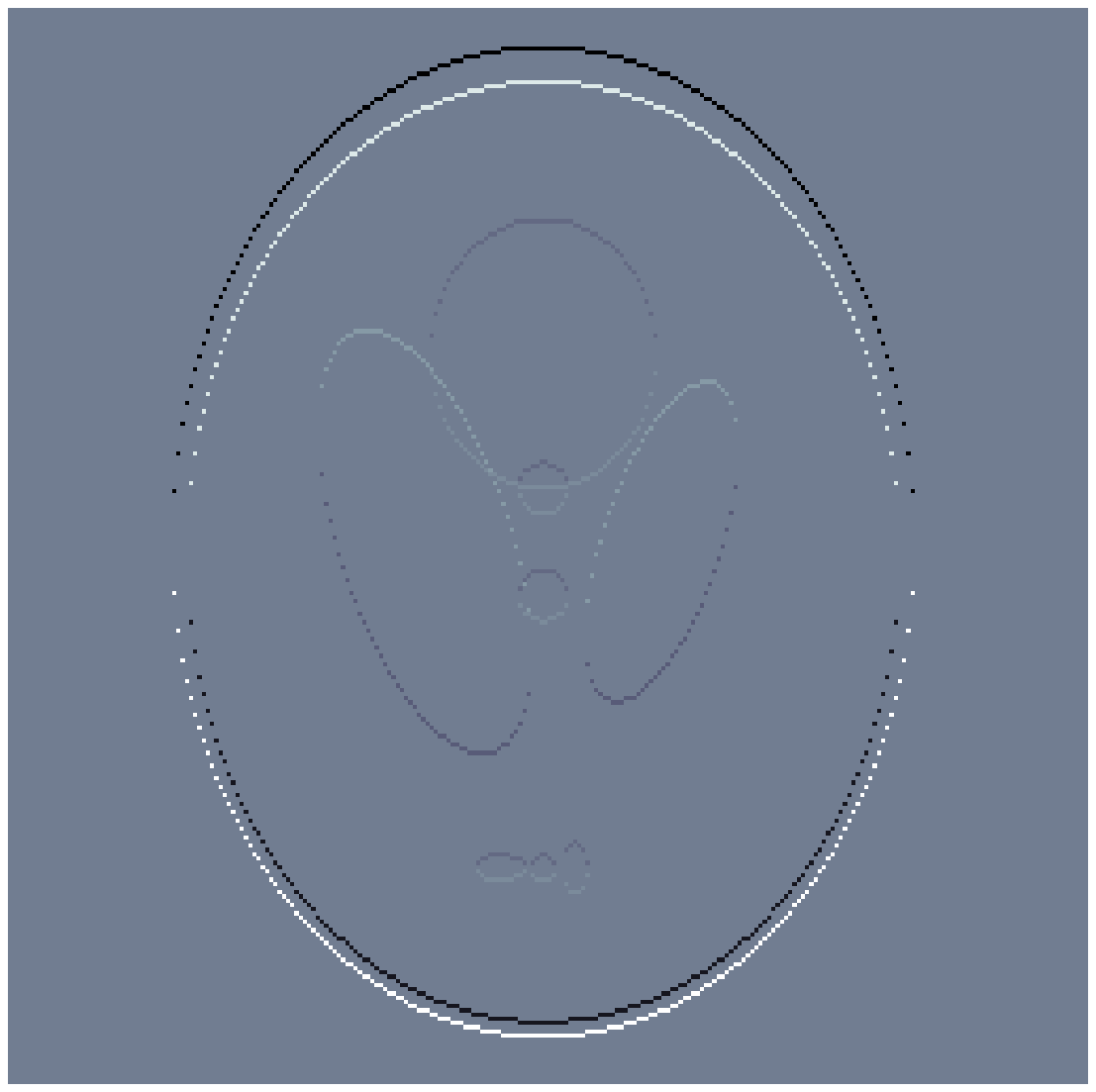}
\includegraphics[width=4.0cm]{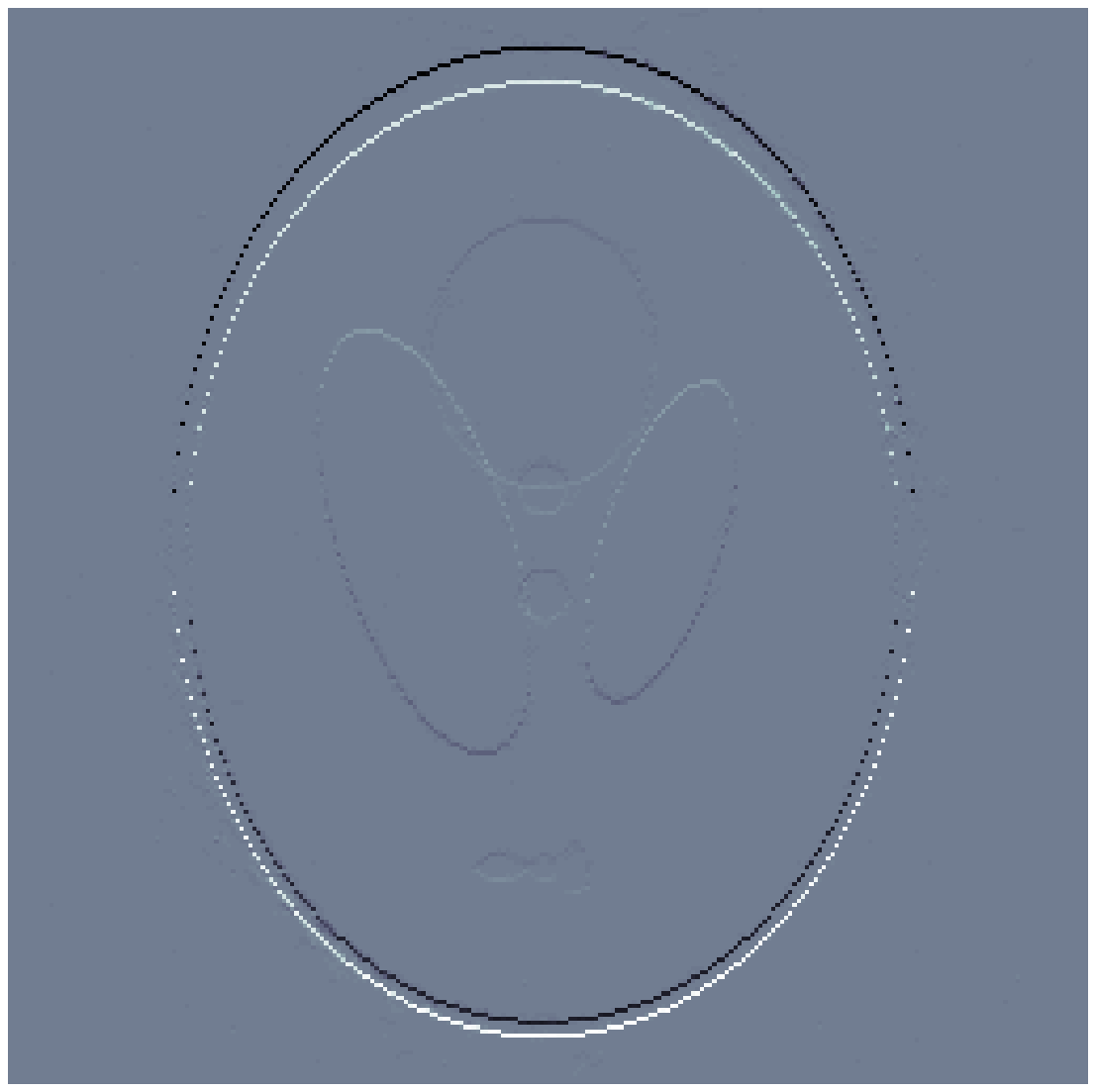}
\caption{Edge detection. (left) $\mathbf{D}_v(\mathbf{x})$ where $\mathbf{x}$ is ground truth. (right) $\mathbf{D}_v(\mathbf{\tilde{x}})$ where $\mathbf{\tilde{x}}$ is obtained via Eq. (\ref{eq:TV}). Note that the $\mathbf{D}_h$ images are also similar but omitted for space.}
\label{fig:edges}
\end{figure}

\begin{figure}[t]
\centering
\includegraphics[width=4.0cm]{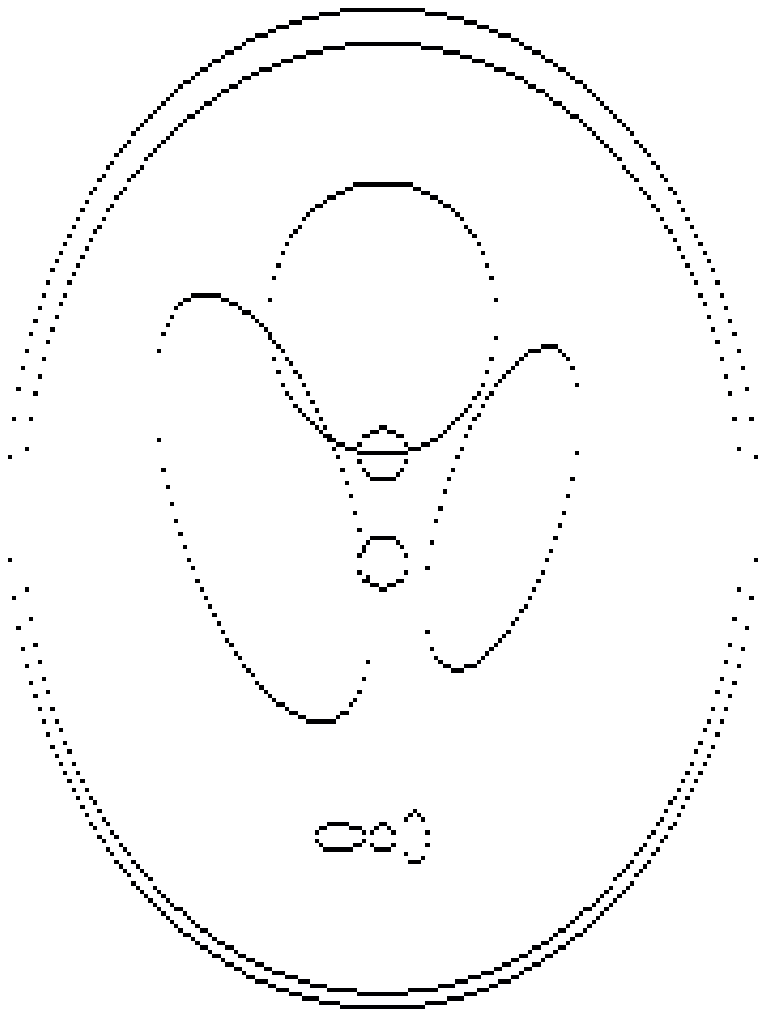}
\includegraphics[width=4.0cm]{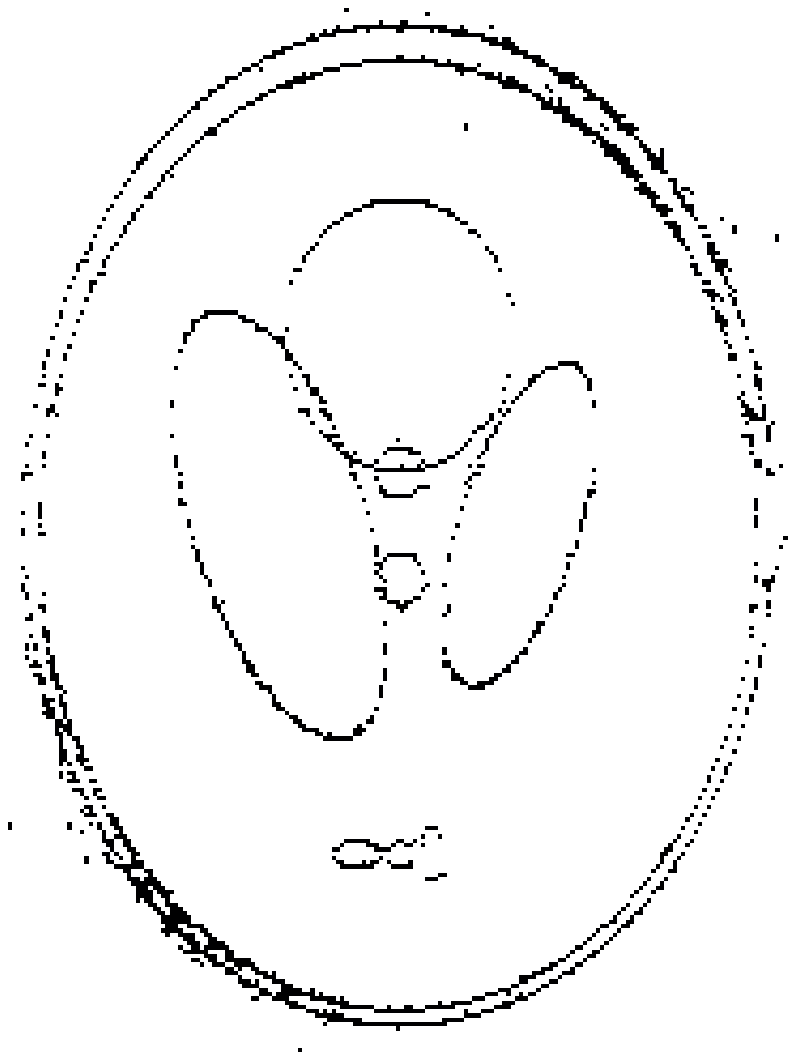}
\caption{Mask creation. (left) $\mathbf{M}_v$ where $\mathbf{x}$ is ground truth. (right) $\mathbf{M}_v$ where $\mathbf{\tilde{x}}$ is obtained via Eq. (\ref{eq:TV}) and $\tau_v(5)$.}
\label{fig:mask}
\end{figure}

To measure accuracy we use the relative error defined by
\begin{align}
RE= \frac{||\mathbf{x}-\mathbf{x}_{true}||_2}{||\mathbf{x}_{true}||_2},
\end{align}
where the difference and norms operate on the vectorized (concatenated) images. For the TV-regularized reconstruction in Figure \ref{fig:TV}, $RE=.0500$. There are visible errors in the intensity values of the image, but the edges appear to be in the proper locations. Figure \ref{fig:edges}, which shows horizontal and vertical edges in the image, confirms this. Specifically, the horizontal and vertical edge transforms are the anisotropic TV transforms defined by
\begin{align}
\left[\mathbf{D}_v(\mathbf{x})\right]_{i,j} &= \sum_k \mathbf{D}_{i,k}\mathbf{x}_{k,j},
\end{align}
and
\begin{align}
\left[\mathbf{D}_h(\mathbf{x})\right]_{i,j} &= \sum_k \mathbf{D}^T_{k,j}\mathbf{x}_{i,k},
\end{align}
where $\mathbf{D}$ is ${N\times N}$ and defined by
\begin{align}\label{eq:TVtransform}
\mathbf{D}_{i,j} = \left\{\begin{array}{cc}
1 & j=i+1\\
-1 & j=i\\
0 & \text{else}\end{array}\right.,
\end{align}
with $\mathbf{D}_{N,1}=1$. Note that there are many other methods for edge detection from image data, including the popular Canny method, \cite{canny1986computational}, which was used in the iteratively reweighted EdgeCS method \cite{guo2010edgecs,guo2012edge}. This paper only considers these anisotropic TV edges. Later the same transforms are used to regularize, ensuring the mask will match the sparsity domain. 

Next, the edge values from Figure \ref{fig:edges} are thresholded to create two binary mask matrices $\mathbf{M}_h$ and $\mathbf{M}_v$, defined by
\begin{align}
\left[\mathbf{M}_v\right]_{i,j} = \left\{\begin{matrix}
1 & |\left[\mathbf{D}_v(\mathbf{x})\right]_{i,j}|<\tau_v\\
0 & |\left[\mathbf{D}_v(\mathbf{x})\right]_{i,j}|\ge\tau_v
\end{matrix}\right.,
\end{align}
and
\begin{align}
\left[\mathbf{M}_h\right]_{i,j} = \left\{\begin{matrix}
1 & |\left[\mathbf{D}_h(\mathbf{x})\right]_{i,j}|<\tau_h\\
0 & |\left[\mathbf{D}_h(\mathbf{x})\right]_{i,j}|\ge\tau_h
\end{matrix}\right..
\end{align}
Figure \ref{fig:mask} shows the exact result as well as the approximate, thresholded edge mask. Similar to \cite{guo2010edgecs,guo2012edge}, the thresholds $\tau_v$ and $\tau_h$ are defined by
\begin{align}\label{eq:thresholds}
\begin{split}
\tau_v(k) &= 2^{-k}\cdot\max\left\{\mathbf{D}_v(\mathbf{\tilde{x}})\right\}\\
\tau_h(k) &= 2^{-k}\cdot\max\{\mathbf{D}_h(\mathbf{\tilde{x}})\}.
\end{split}
\end{align}
where $k$ is set by the user. For example, choosing $k=5$ marks all grid points above $3.125\%$ of the maximum edge value as edges. In general $k$ is a resolution and noise dependent parameter.

\begin{figure}[t]
\centering
\includegraphics[width=4.0cm]{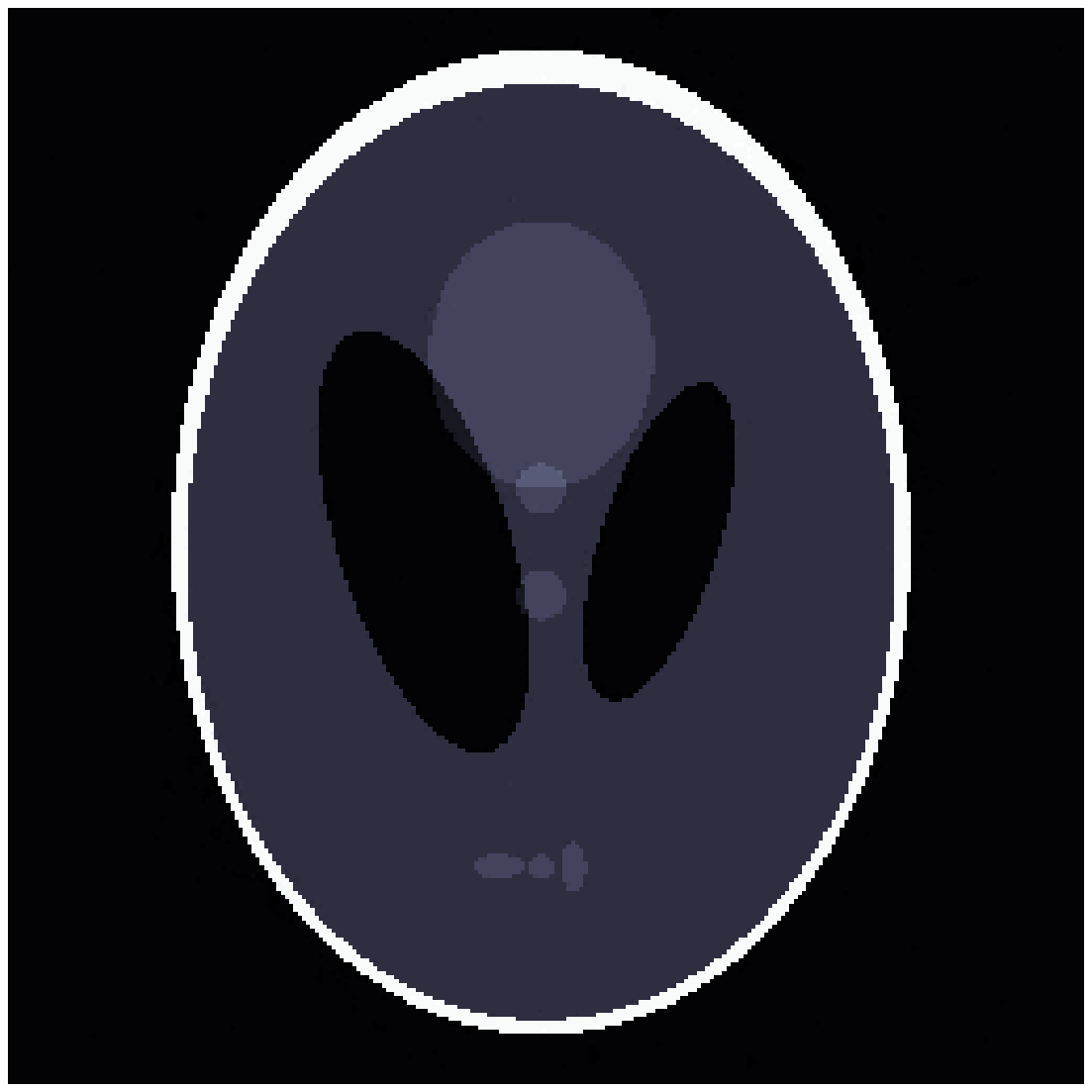}
\includegraphics[width=4.0cm]{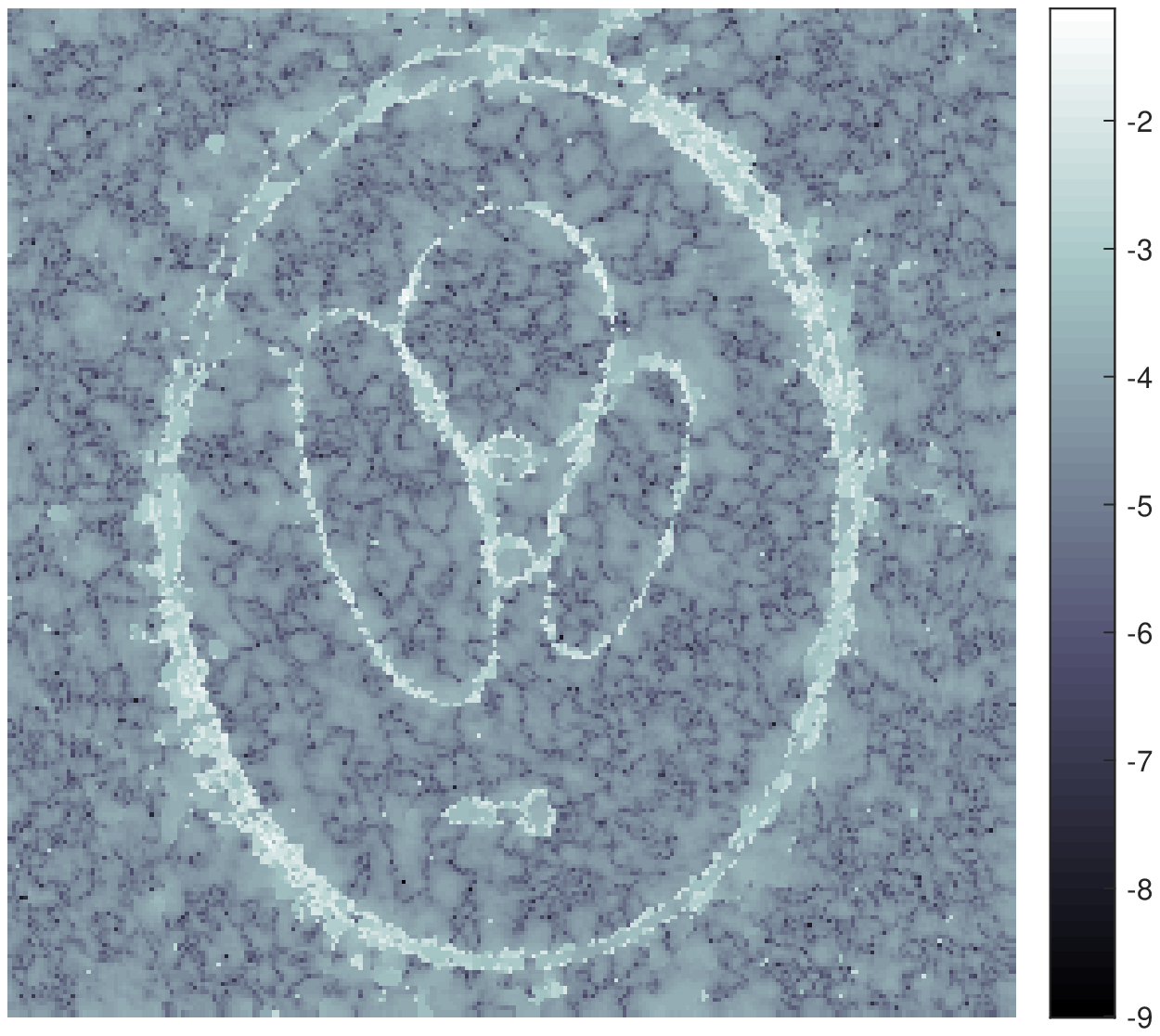}
\includegraphics[width=8.0cm]{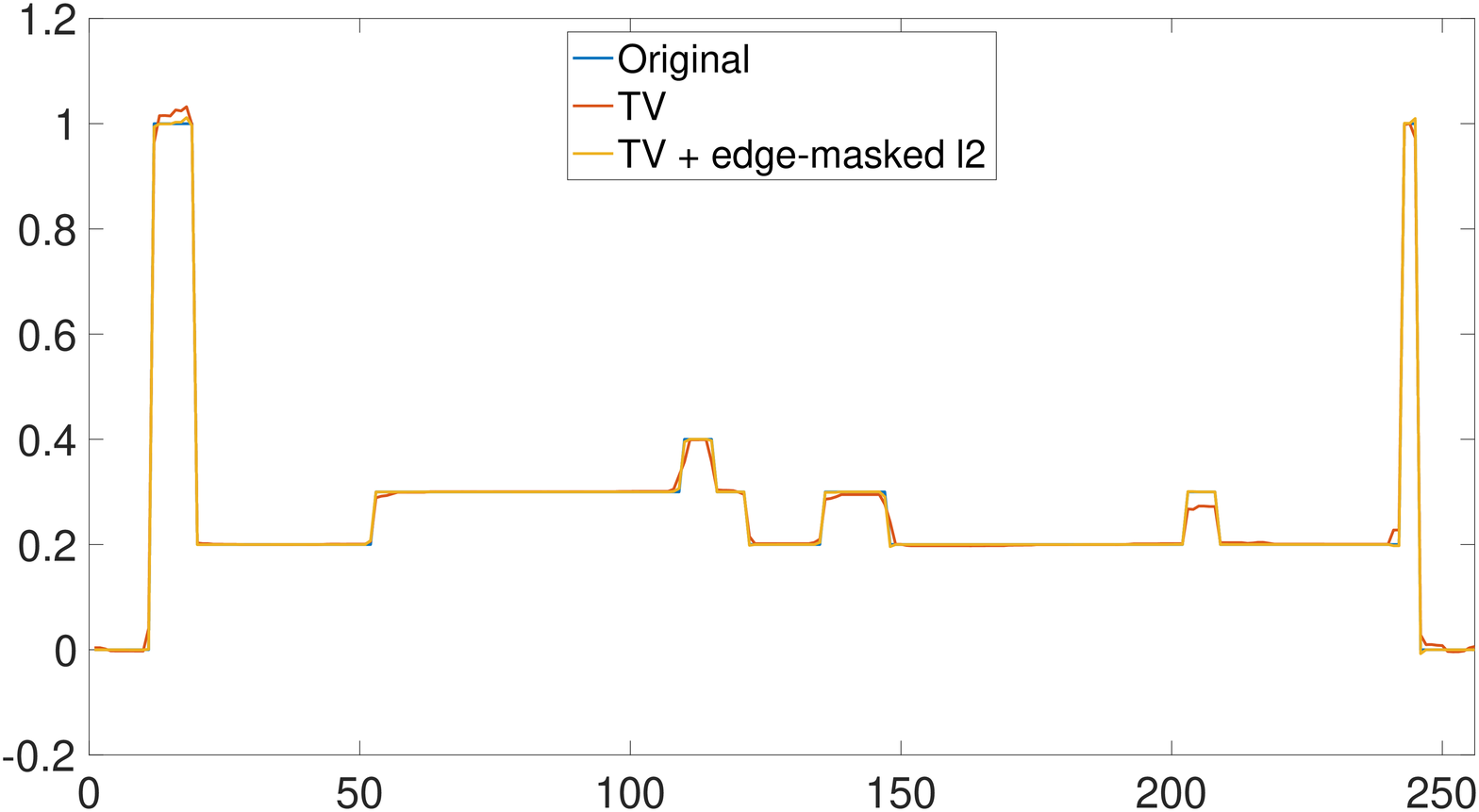}
\caption{Edge-masked enhancement of Eq. (\ref{eq:TV}) result from 16 radial lines with point-wise error plot and vertical cross-section.}
\label{fig:Xadapt}
\end{figure}

\begin{table}
\begin{center}
 \begin{tabular}{|c|c|c|c|} 
 \hline
  Radial lines & TV & TV + edge-masked $\ell_2$ &$k$\\
 \hline
 16 & .0500 & .0063 & 256 \\ 
 \hline
 15 & .0769 & .0159 & 64 \\
 \hline
 14 & .1246 & .0330 & 32 \\
 \hline
 13 & .1763 & .0518 & 32 \\
 \hline
 12 & .3189 & .1779 & 32 \\ 
 \hline
\end{tabular}
\caption{Relative errors for TV and TV plus edge-masked $\ell_2$ enhancement using radial line data and thresholds defined by $k$ in Eq. (\ref{eq:thresholds}).}
\end{center}
\end{table}\label{table:1}

Finally, we employ the two masks to perform the image enhancement via the reconstruction
\begin{align}\label{eq:EM}
\begin{split}
&\arg\min_\mathbf{x} \left|\left| \begin{bmatrix}
\mathbf{M}_v\odot\mathbf{D}_v(\mathbf{x})\\
\mathbf{M}_h\odot\mathbf{D}_h(\mathbf{x})\end{bmatrix}\right|\right|_2^2 
\text{subject to}\quad \mathbf{Fx}=\mathbf{\hat{f}}.
\end{split}
\end{align}
Here $\odot$ denotes elementwise multiplication. When noise is present, Eq. (\ref{eq:EM}) is modified to
\begin{align}\label{eq:EMnoisy}
\begin{split}
&\arg\min_\mathbf{x} ||\mathbf{Fx}-\mathbf{\hat{f}}||_2^2+\lambda\left|\left| \begin{bmatrix}
\mathbf{M}_v\odot\mathbf{D}_v(\mathbf{x})\\
\mathbf{M}_h\odot\mathbf{D}_h(\mathbf{x})\end{bmatrix}\right|\right|_2^2.
\end{split}
\end{align}
The anisotropic TV formulation is used for regularization as it was shown in \cite{guo2010edgecs,guo2012edge} to be more effective in an edge-weighting scheme than isotropic TV. Empirical and theoretical evidence from \cite{churchill2018edge,churchill2019edge} also supports the use of anisotropic TV. Figure \ref{fig:Xadapt} shows the result for the Shepp-Logan phantom reconstructed from 16 radial lines. The relative error is reduced from $.0500$ to $.0063$, and the error plot shows drastic accuracy improvement particularly in smooth regions. Similar results were achieved when supplementing TV-regularized reconstruction from 12, 13, 14, and 15 radial lines with edge-masked regularization as well. Table \ref{table:1} compares the relative errors for both methods.

\section{Results}\label{sec:results}

\begin{figure}[t]
\centering
\includegraphics[width=4.0cm]{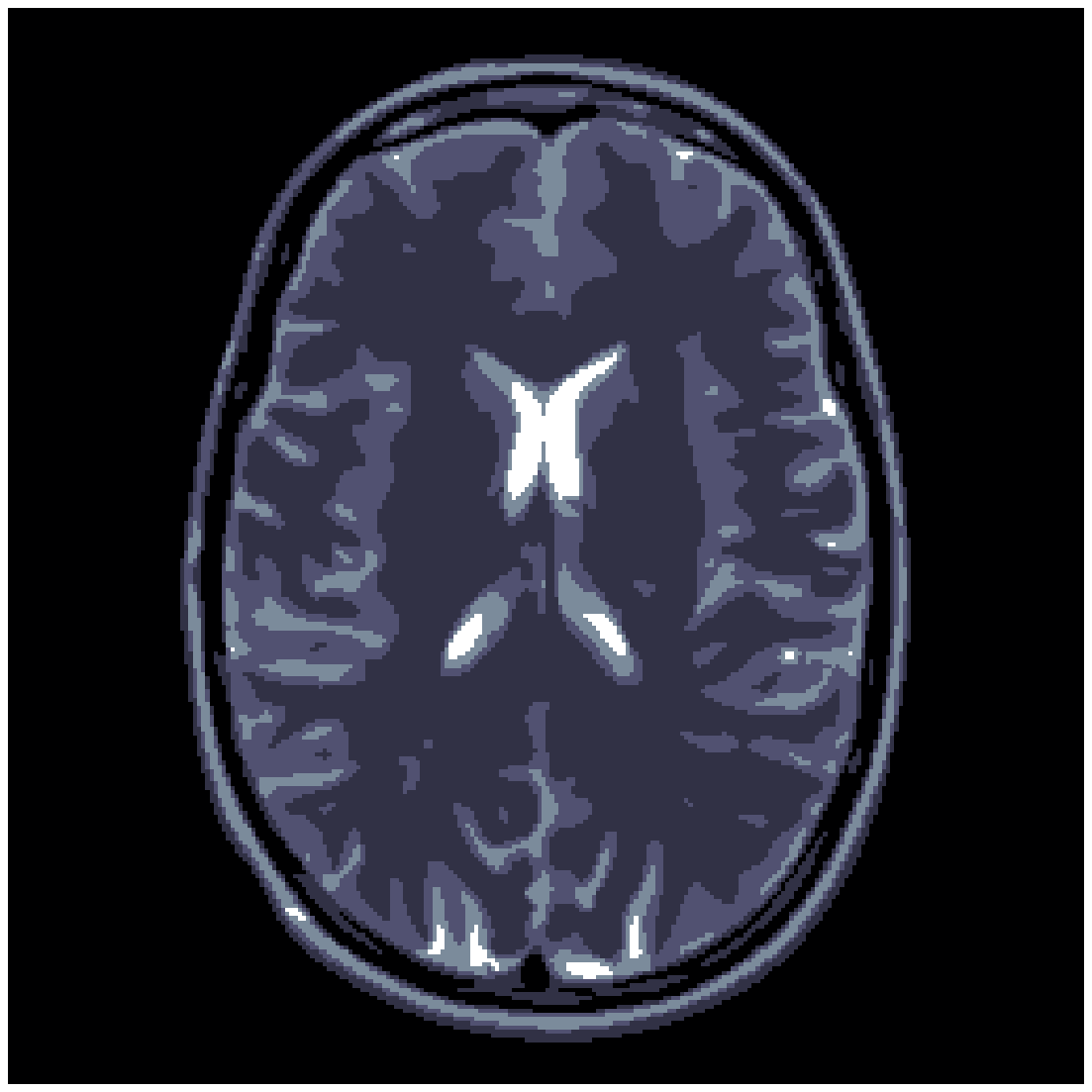}
\includegraphics[width=4.0cm]{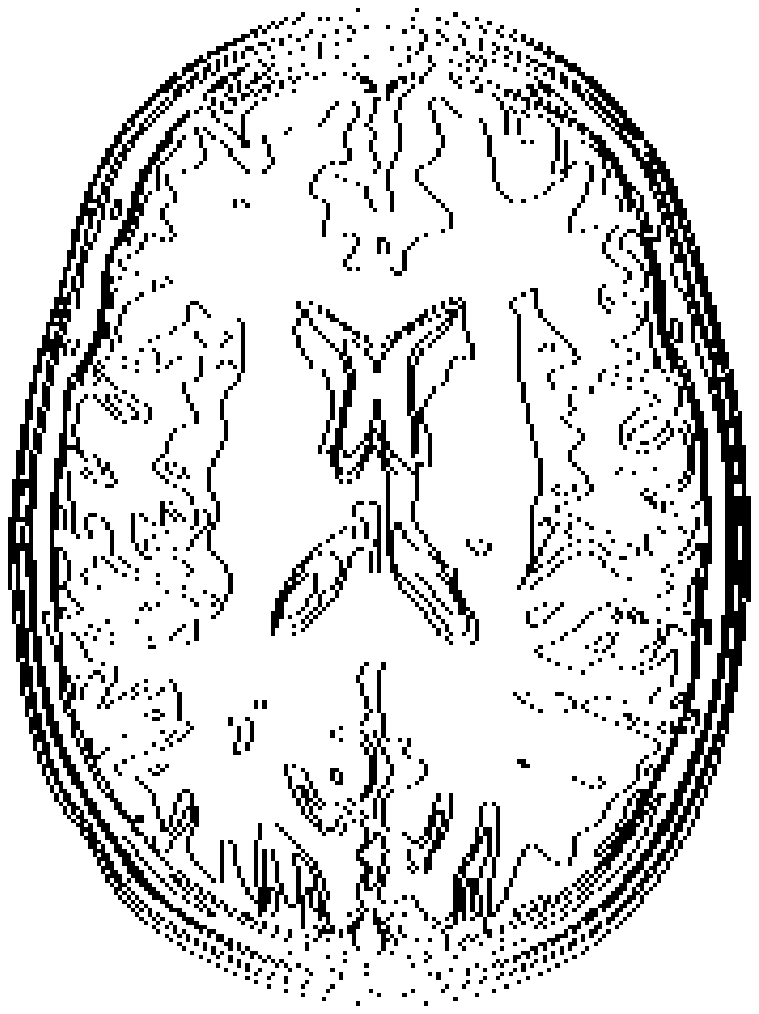}
\caption{(left) Realistic brain phantom \cite{guerquin2012realistic} and (right) true vertical edge mask.}
\label{fig:mri}
\end{figure}

Motivated by magnetic resonance (MR) imaging, where Fourier data is often collected along radial lines, the following experiments use a realistic brain phantom from \cite{guerquin2012realistic}, shown in the left panel of Figure \ref{fig:mri}. This image is much more difficult to reconstruct, e.g. requiring 77 radial lines of Fourier data to achieve relative error of less than $10^{-2}$ in the noise-less case using isotropic TV regularization as in Eq. (\ref{eq:TV}). This is due to its overall higher total variation and dense edge structure, seen in the right panel of Figure \ref{fig:mri}, compared with the Shepp-Logan phantom.

\begin{figure}[t]
\centering
\includegraphics[width=4.0cm]{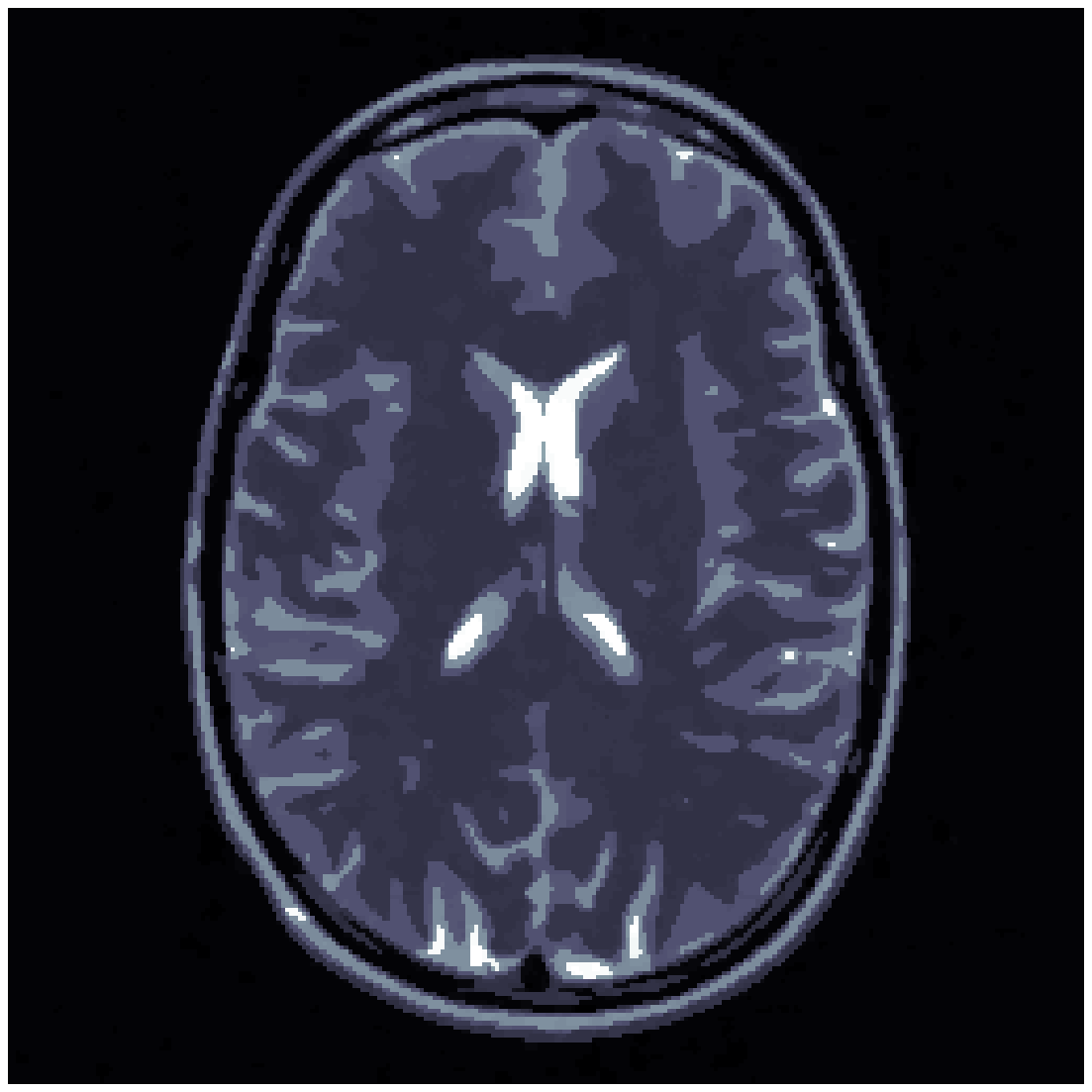}
\includegraphics[width=4.0cm]{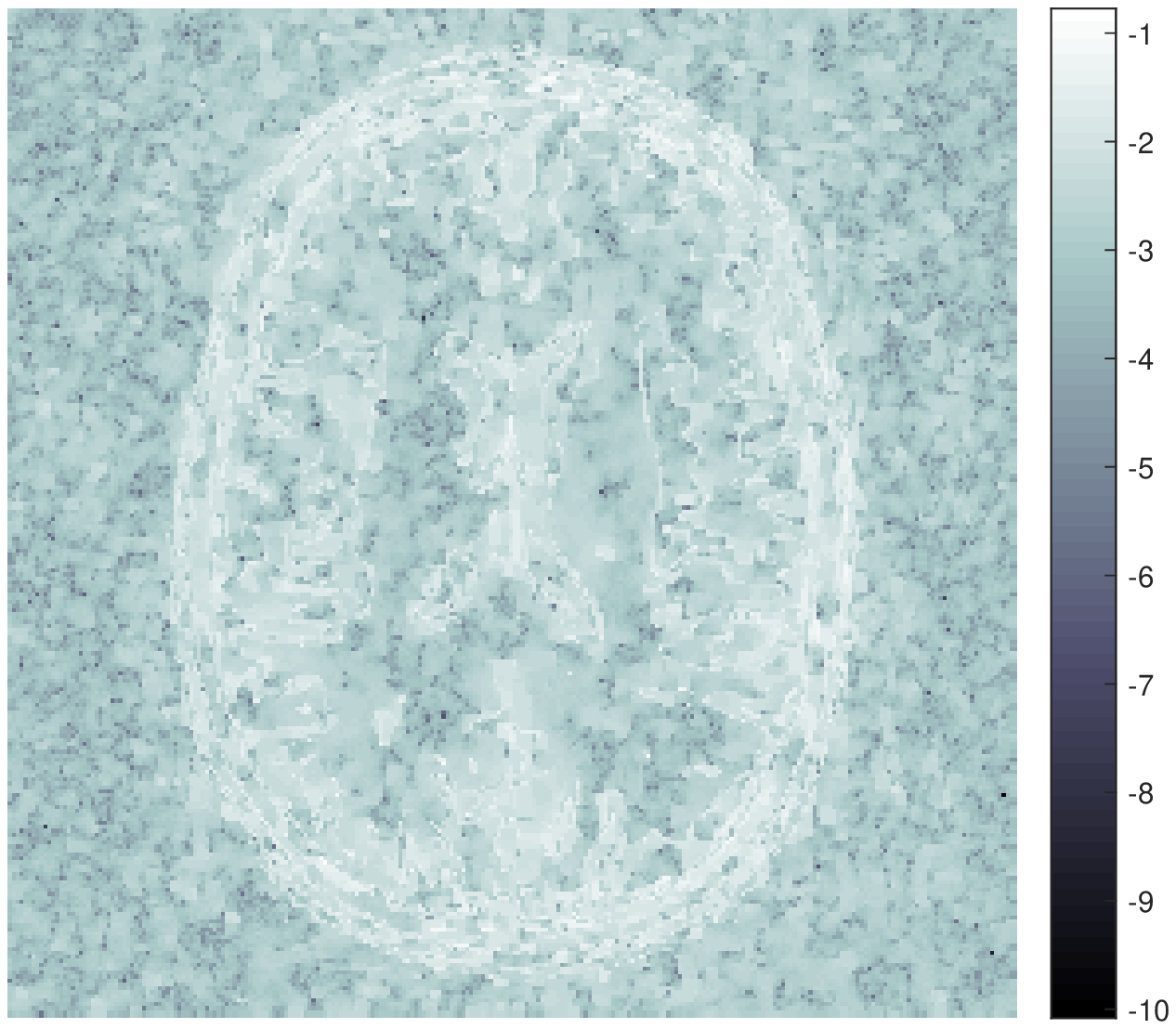}
\includegraphics[width=4.0cm]{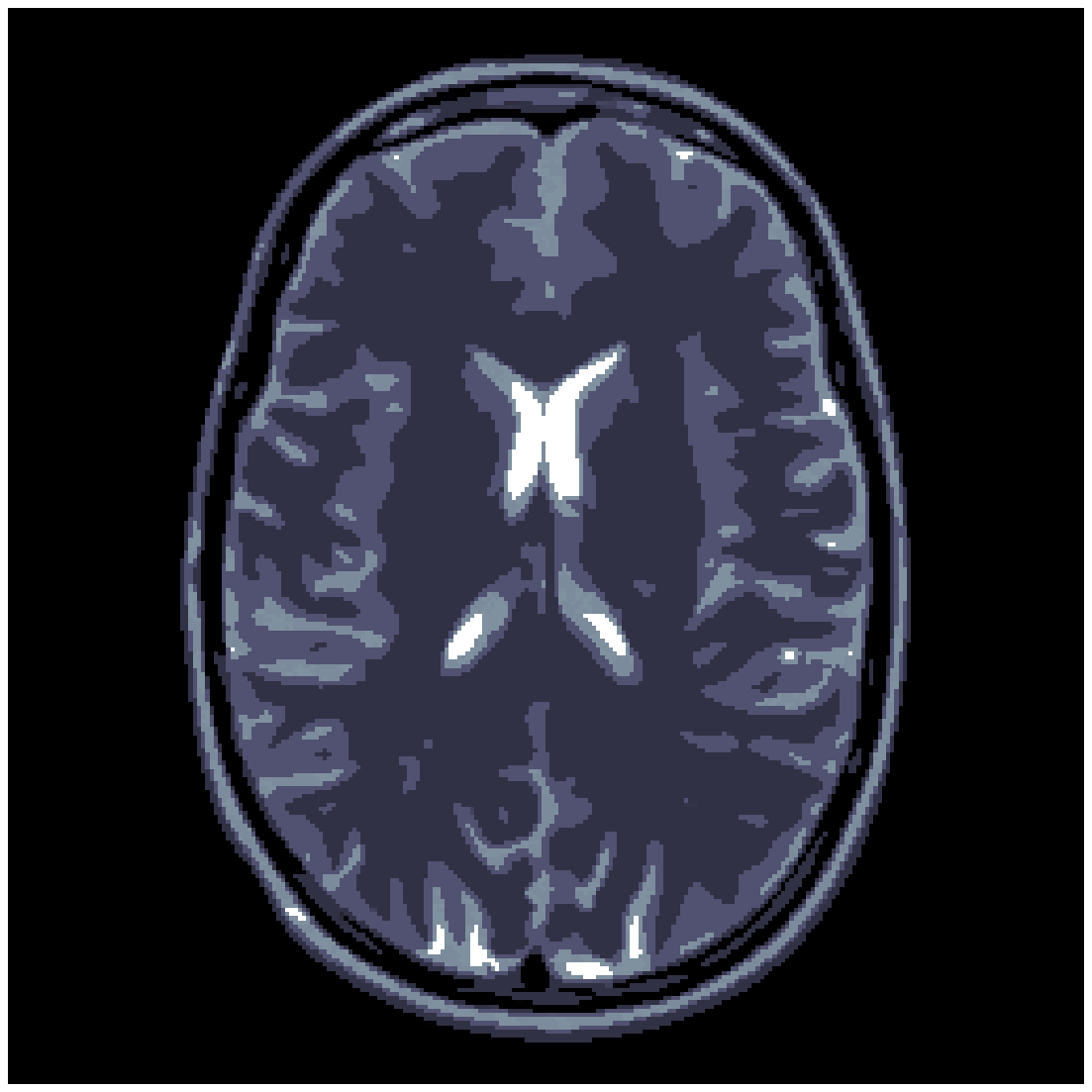}
\includegraphics[width=4.0cm]{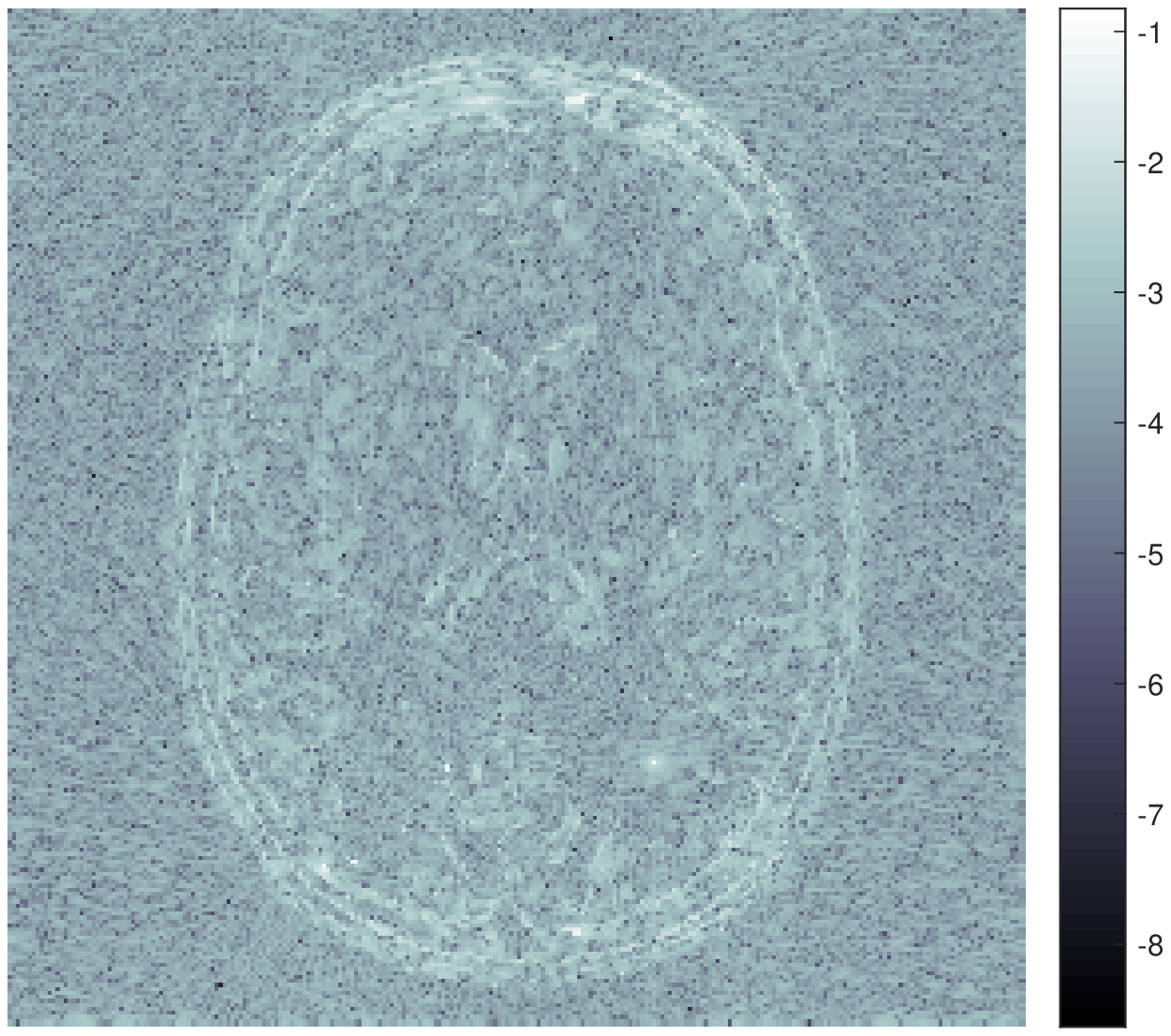}
\includegraphics[width=8.0cm]{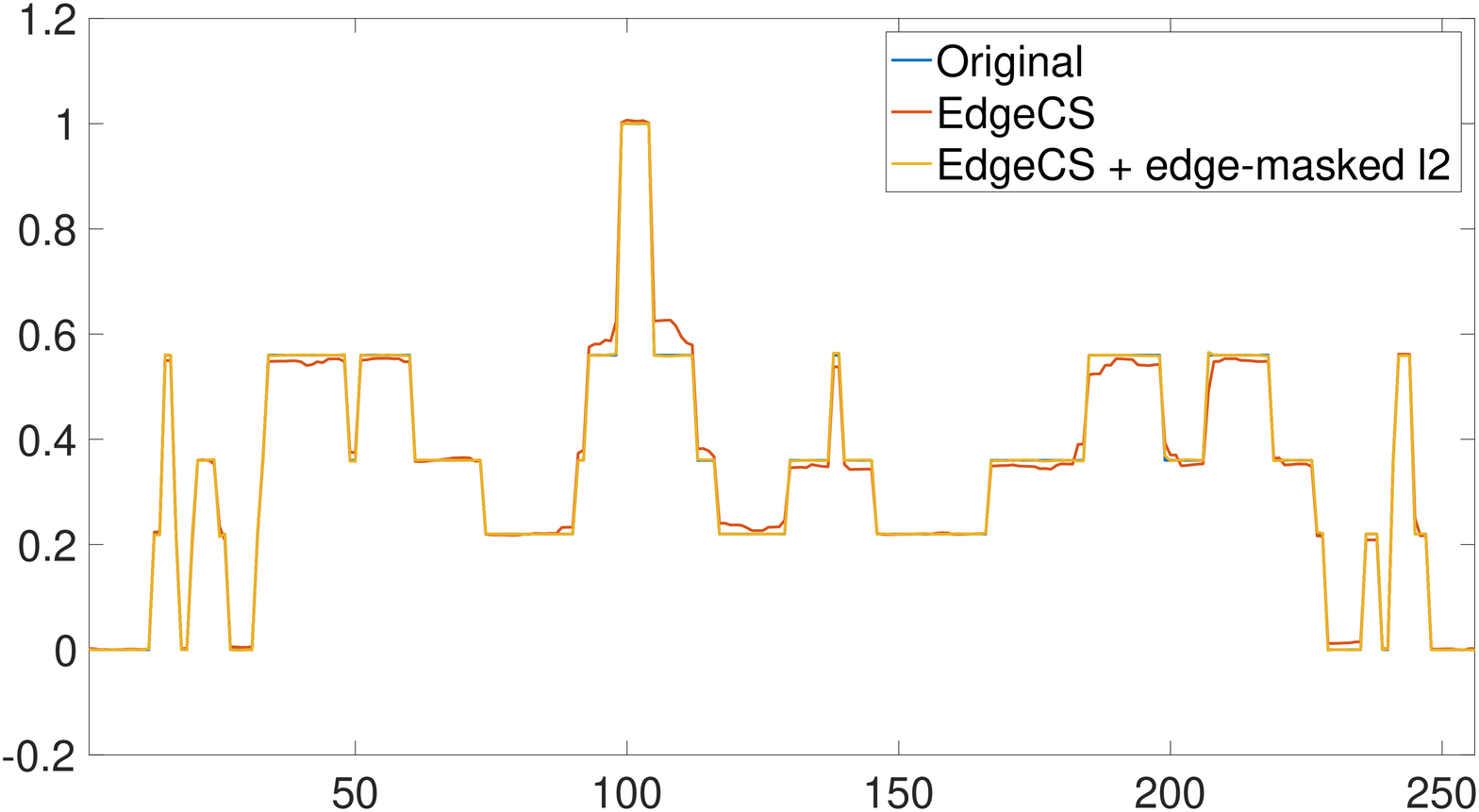}
\caption{Reconstruction and enhancement from limited data. (top) EdgeCS image from 34 radial lines with point-wise error. (middle) edge-masked enhancement with point-wise error. (bottom) vertical cross-section comparison. Here $k=5$.}
\label{fig:mrilimited}
\end{figure}

\begin{figure}[t]
\centering
\includegraphics[width=4.0cm]{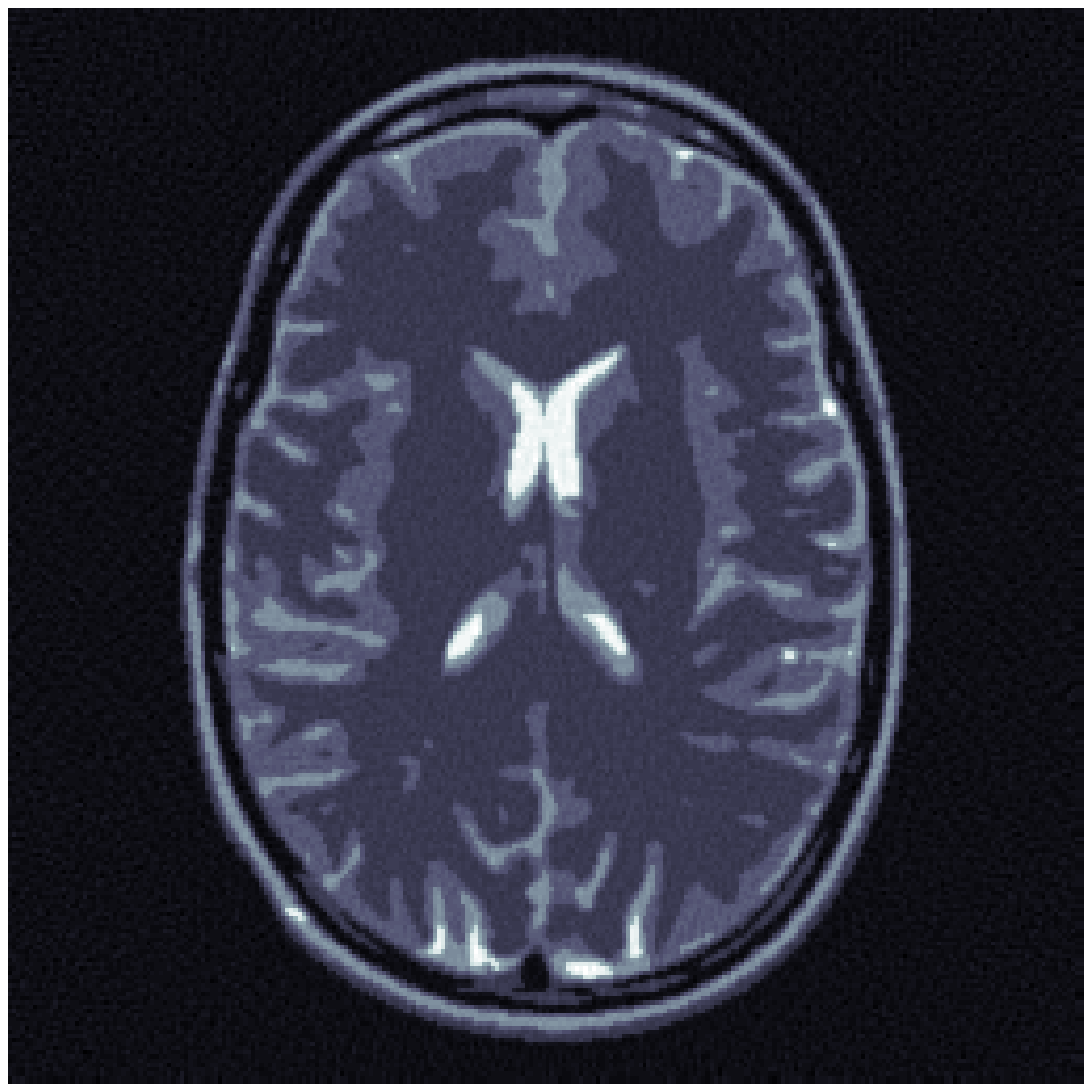}
\includegraphics[width=4.0cm]{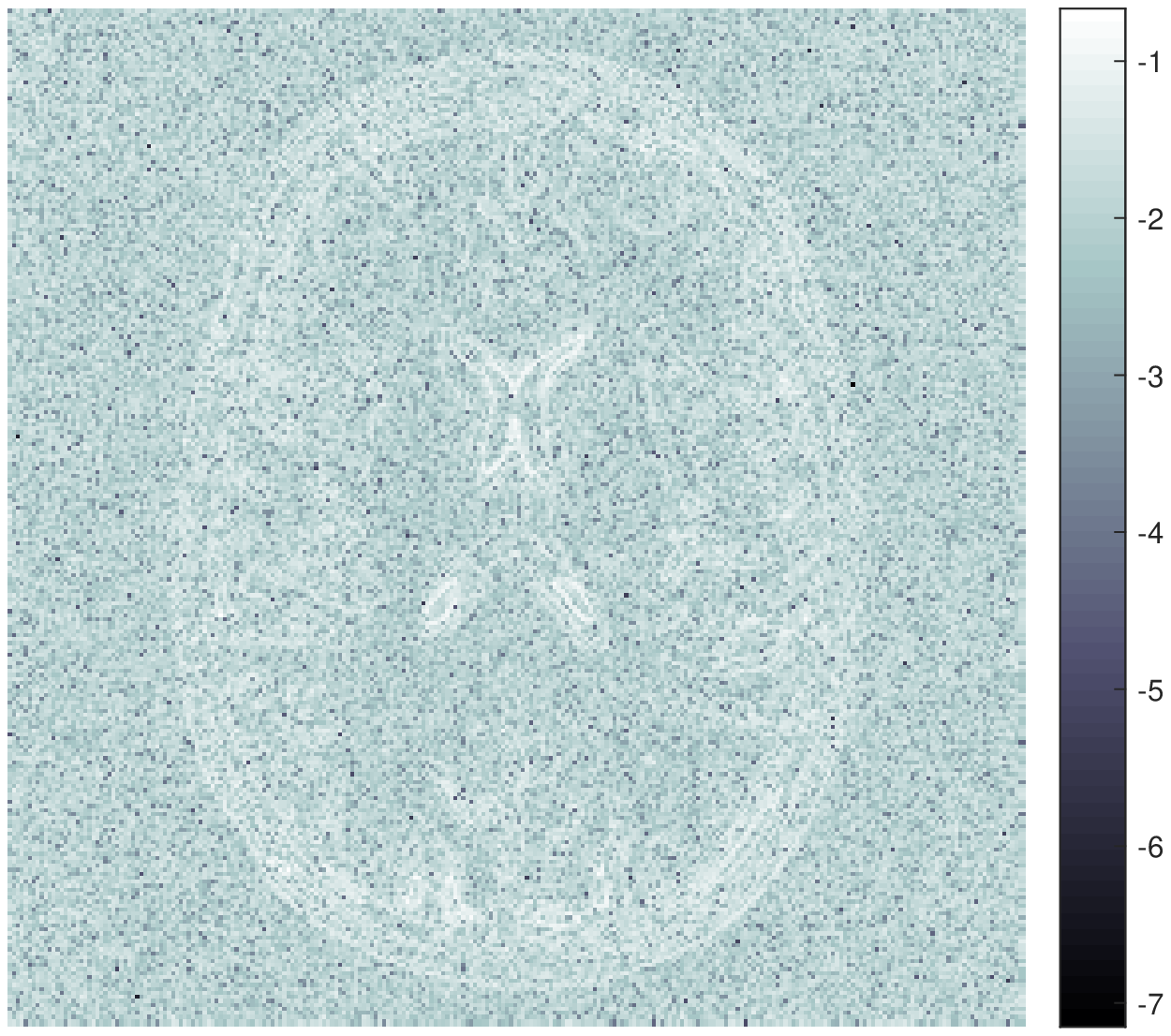}
\includegraphics[width=4.0cm]{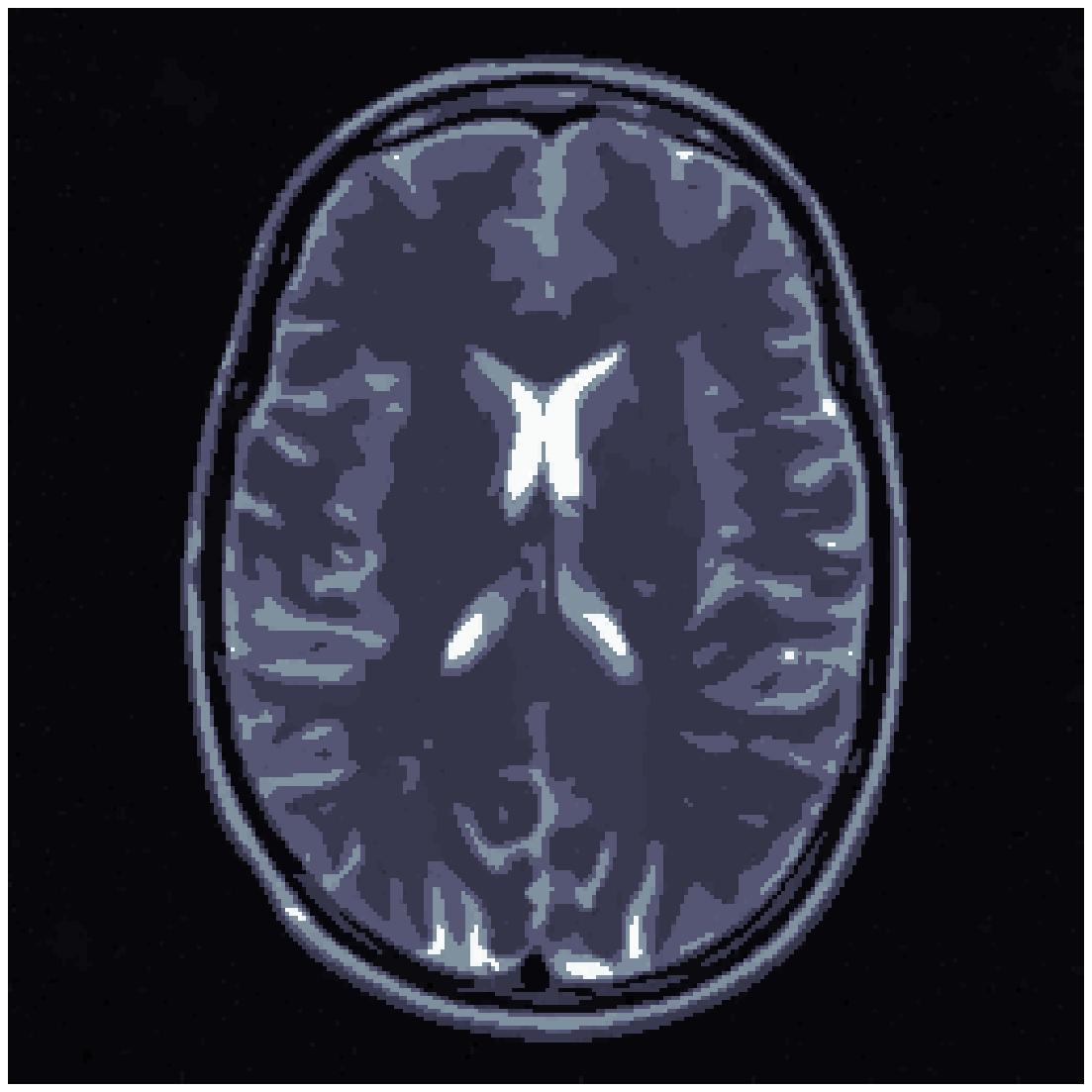}
\includegraphics[width=4.0cm]{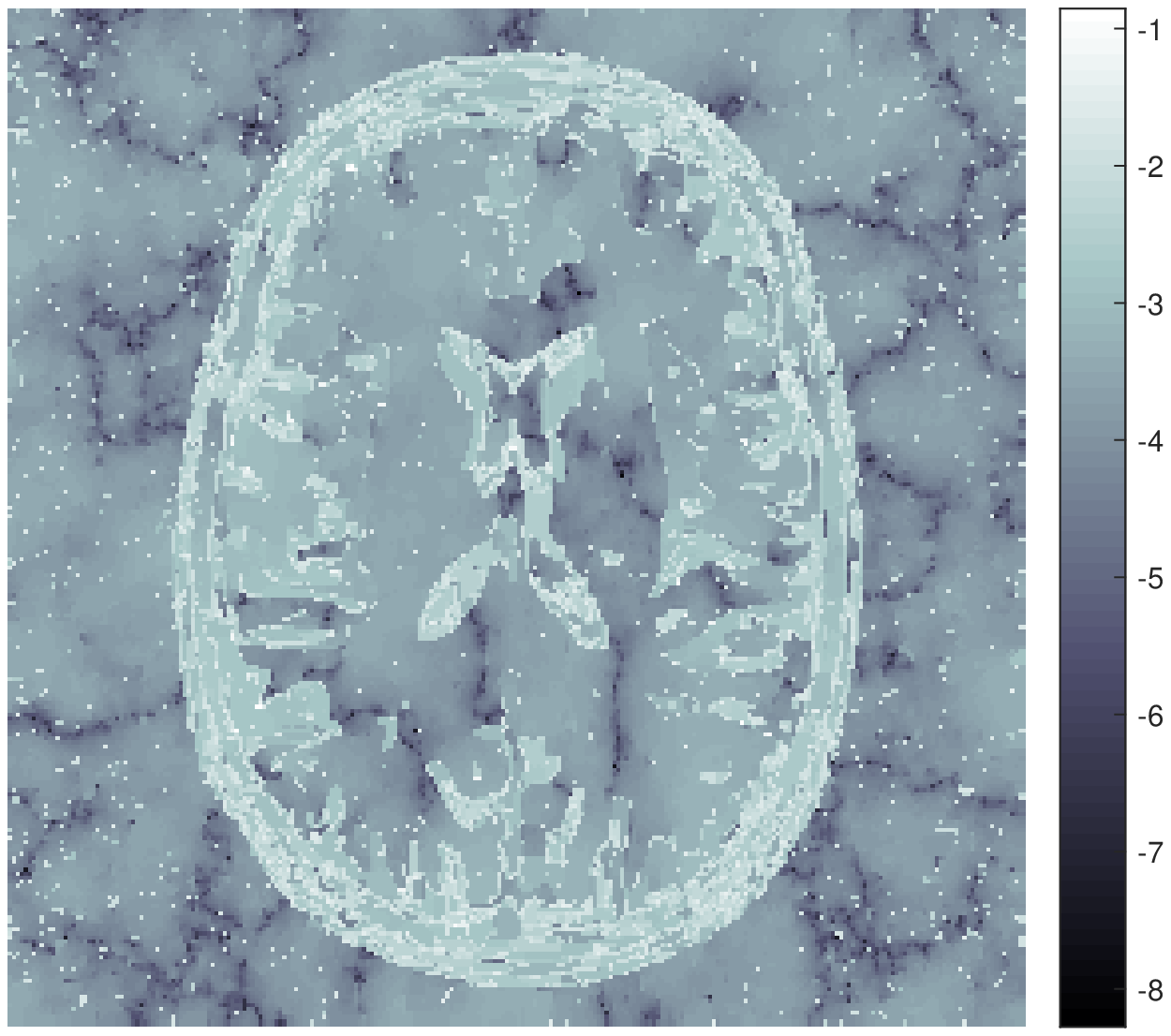}
\includegraphics[width=8.0cm]{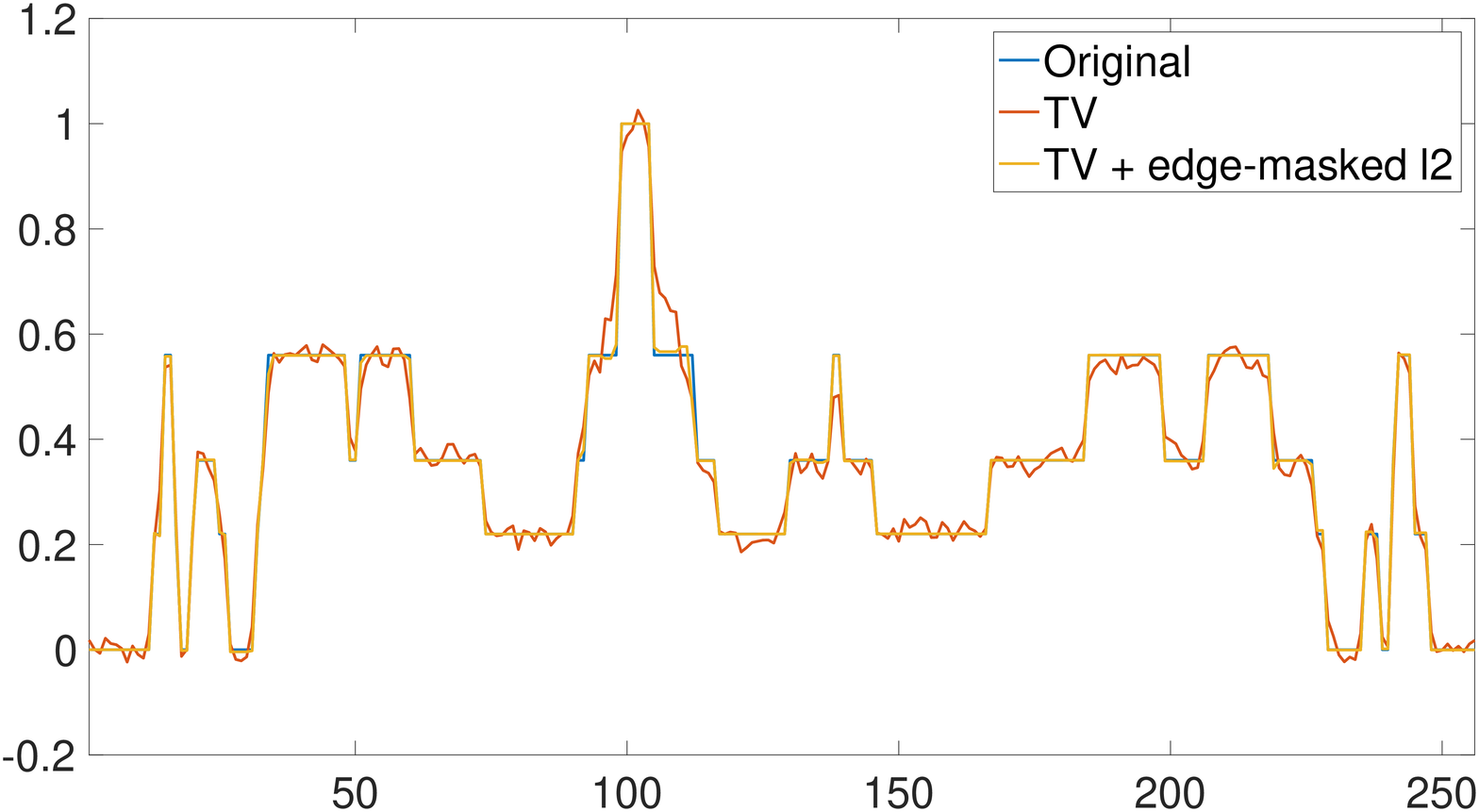}
\caption{Reconstruction and enhancement from noisy data. (top) ``un-masked'' anisotropic TV image from 180 radial lines with added zero-mean Gaussian noise with standard deviation $10^{-2}$ with point-wise error. (middle) edge-masked enhancement with point-wise error. (bottom) vertical cross-section comparison. The initial reconstruction uses $\lambda=10^{-9}$ and the enhancement uses $\mu=10^{-2}$, and $k=4$.}
\label{fig:mrinoisy}
\end{figure}

\subsection{Very limited data}
In this experiment, the initial reconstruction is via EdgeCS \cite{guo2010edgecs,guo2012edge}, an iteratively edge-weighted reconstruction method. EdgeCS requires only 36 radial lines to near-perfectly reconstruct the phantom. For the initial reconstruction in this experiment, 34 lines of data are used, and the relative error is $.0344$. The edge-masked enhancement step improves it to $.0072$, as Figure \ref{fig:mrilimited} shows. This experiment provides some evidence that even with very limited data and an advanced initial reconstruction, there is still room for the edge-masked regularization to improve accuracy and enhance edge-sparsity.

\subsection{Additive noise}
In this experiment an ``un-masked'' Eq. (\ref{eq:EMnoisy}), i.e. $\mathbf{M}_h$ and $\mathbf{M}_v$ all ones, is used to initially reconstruct the phantom from 180 radial lines of data with zero-mean Gaussian noise with standard deviation $10^{-2}$ added. The image has a relative error of $.0947$. The edge-masked enhancement achieves a relative error of $.0219$. Figure \ref{fig:mrinoisy} shows the result. Note that the faithfulness of the initial reconstruction to the true edge locations here is paramount as the enhancement especially relies on an accurate edge mask when noise is present. 

\section{Conclusion}\label{sec:conclusion}
This paper presented an algorithm to enhance images reconstructed via edge-sparsity based methods when data requirements for near-perfect reconstruction are not met. It is able to achieve this because while the intensity values in the resulting images may not be ideal, the edge locations are often faithful to those of the ground truth. The algorithm locates the edges and uses them in a masked $\ell_2$ regularization scheme. Our method was shown to further enhance edge information and improve accuracy for three different initial reconstruction methods, varying amounts of limited data with and without noise, and two different phantom images. In future work, in order to further boost our enhancement results, we will explore edge-sparsity based methods that are robust with respect to noise for use in our initial reconstruction step. In addition, we will explore an iteratively ``re-masked'' algorithm similar to \cite{guo2010edgecs,guo2012edge} but using $\ell_2$ regularization instead of $\ell_1$.

\bibliographystyle{IEEEbib}
\bibliography{refs}

\begin{thebibliography}{10}

\bibitem{candes2006robust}
Emmanuel~J Cand{\`e}s, Justin Romberg, and Terence Tao,
\newblock ``Robust uncertainty principles: Exact signal reconstruction from
  highly incomplete frequency information,''
\newblock {\em IEEE Transactions on information theory}, vol. 52, no. 2, pp.
  489--509, 2006.

\bibitem{churchill2018edge}
Victor Churchill, Rick Archibald, and Anne Gelb,
\newblock ``Edge-adaptive $\ell_2$ regularization image reconstruction from
  non-uniform {F}ourier data,''
\newblock {\em arXiv preprint arXiv:1811.08487}, 2018,
\newblock https://arxiv.org/abs/1811.08487.

\bibitem{churchill2019edge}
Victor Churchill and Anne Gelb,
\newblock ``Edge-masked {CT} image reconstruction from limited data,''
\newblock {\em arXiv preprint arXiv:1901.05275}, 2019,
\newblock https://arxiv.org/abs/1901.05275.

\bibitem{candes2008enhancing}
Emmanuel~J Candes, Michael~B Wakin, and Stephen~P Boyd,
\newblock ``Enhancing sparsity by reweighted $\ell_1$ minimization,''
\newblock {\em Journal of Fourier analysis and applications}, vol. 14, no. 5,
  pp. 877--905, 2008.

\bibitem{chartrand2008iteratively}
Rick Chartrand and Wotao Yin,
\newblock ``Iteratively reweighted algorithms for compressive sensing,''
\newblock in {\em Acoustics, Speech and Signal Processing, 2008. ICASSP 2008.
  IEEE International Conference on}. IEEE, 2008, pp. 3869--3872.

  \vfill
  \pagebreak

\bibitem{guo2010edgecs}
Weihong Guo and Wotao Yin,
\newblock ``Edge{CS}: Edge guided compressive sensing reconstruction,''
\newblock in {\em Visual Communications and Image Processing 2010}.
  International Society for Optics and Photonics, 2010, vol. 7744, p. 77440L.
  


\bibitem{guo2012edge}
Weihong Guo and Wotao Yin,
\newblock ``Edge guided reconstruction for compressive imaging,''
\newblock {\em SIAM Journal on Imaging Sciences}, vol. 5, no. 3, pp. 809--834,
  2012.

\bibitem{rudin1992nonlinear}
Leonid~I Rudin, Stanley Osher, and Emad Fatemi,
\newblock ``Nonlinear total variation based noise removal algorithms,''
\newblock {\em Physica D: Nonlinear Phenomena}, vol. 60, no. 1-4, pp. 259--268,
  1992.

\bibitem{shepp1974fourier}
Lawrence~A Shepp and Benjamin~F Logan,
\newblock ``The {F}ourier reconstruction of a head section,''
\newblock {\em IEEE Transactions on nuclear science}, vol. 21, no. 3, pp.
  21--43, 1974.

\bibitem{canny1986computational}
John Canny,
\newblock ``A computational approach to edge detection,''
\newblock {\em IEEE Transactions on pattern analysis and machine intelligence},
  , no. 6, pp. 679--698, 1986.

\bibitem{guerquin2012realistic}
Matthieu Guerquin-Kern, Laurent Lejeune, Klaas~Paul Pruessmann, and Michael
  Unser,
\newblock ``Realistic analytical phantoms for parallel magnetic resonance
  imaging,''
\newblock {\em IEEE Transactions on Medical Imaging}, vol. 31, no. 3, pp.
  626--636, 2012.

\end{thebibliography}



\end{document}